\documentclass[aps,prb,twocolumn,showpacs,superscriptaddress]{revtex4-1}
\usepackage{amssymb,amsmath}
\usepackage{graphicx}
\usepackage{dcolumn}
\usepackage{bm}
\usepackage[normalem]{ulem}
\usepackage{color}

\begin{document}

\title{First-Principles Modeling of Quantum Nuclear Effects and Atomic 
             Interactions in Solid $^{\bf 4}$He at High Pressure}

\author{Claudio Cazorla}
\affiliation{School of Materials Science and Engineering,
             University of New South Wales, Sydney NSW 2052, Australia}
\author{Jordi Boronat}
\affiliation{Departament de F\'{i}sica i Enginyeria Nuclear, 
             Universitat Polit\`{e}cnica de Catalunya, Campus 
             Nord B4-B5, E-08034, Barcelona, Spain}
\email{ccazorla@icmab.es}

\begin{abstract}
We present a first-principles computational study of solid $^{4}$He 
at $T = 0$~K and pressures up to $\sim 160$~GPa. Our computational strategy
consists in using van der Waals density functional theory (DFT-vdW) to describe 
the electronic degrees of freedom in this material, and the diffusion Monte 
Carlo (DMC) method to solve the Schr\"odinger equation describing the behavior 
of the quantum nuclei. For this, we construct an analytical interaction function 
based on the pairwise Aziz potential that closely matches the volume variation of 
the cohesive energy calculated with DFT-vdW in dense helium. Interestingly, we 
find that the kinetic energy of solid $^{4}$He does not increase appreciably with 
compression for $P \ge 85$~GPa. Also, we show that the Lindemann ratio in dense
solid $^{4}$He amounts to $0.10$ almost independently of pressure. The reliability 
of customary quasi-harmonic DFT (QH DFT) approaches in the description of quantum 
nuclear effects in solids is also studied. We find that QH DFT simulations, although 
provide a reasonable equation of state in agreement with experiments, are not able to 
reproduce correctly these critical effects in compressed $^{4}$He. In particular, 
we disclose huge discrepancies of at least $\sim 50$~\% in the calculated $^{4}$He 
kinetic energies using both the QH DFT and present DFT-DMC methods. 
\end{abstract}
\maketitle

\section{Introduction}
\label{sec:introduction}
Solid helium typifies an extreme quantum condensed-matter system. Due to the 
light mass of the atoms and weak interparticle interactions, quantum nuclear 
delocalization effects become crucially important in this crystal. At 
absolute zero temperature $^{4}$He atoms move agitatedly around the equilibrium 
positions of their hexagonal closed packed (hcp) lattice, producing unusually 
large kinetic energies (that is, comparable in magnitude to the potential energy),
and major anharmonic effects.~\cite{moleko85,diallo04,burns97} 
Yet, it has been debated, based on the observations of non-classical rotational
inertia phenomena, that $^{4}$He crystals could behave partly as a fluid with
zero viscosity.~\cite{chan04a,chan04b,cazorla06,cazorla09,rittner07,chan12,day07,rojas10}

In order to fully understand and make quantitative predictions on the quantum 
nature of solid $^{4}$He, it is necessary to solve the corresponding master equations 
of quantum mechanics. This represents an extremely challenging mathematical 
problem due to the non-linearity of the equations involved and large number of 
nuclear and electronic degrees of freedom to be considered. Fortunately, at 
normal conditions helium atoms are, from an electronic band-structure point 
of view, very elementary particles thereby the $^{4}$He--$^{4}$He interactions can 
be effectively modeled with simple analytical expressions that exclusively depend on 
the interatomic distances (e.g., Lennard-Jones and Aziz like potentials).~\cite{aziz87,boronat94} 
By making use of these simplifications and employing advanced quantum simulation 
methods (e.g., quantum Monte Carlo), it has been possible 
to determine with tremendous accuracy and computational efficiency the ground-state 
properties of solid $^4$He.~\cite{boronat94,ceperley95} 
The same kind of approach has been successfully applied also to the study of similar 
systems like H$_{2}$, LiH, LiD, and Ne.~\cite{ceperley87,cazorla08b,cazorla04,cazorla05,cazorla08d} 

A fundamental question that remains to be answered at the quantitative level 
is: how important quantum nuclear effects turn out to be in solid helium 
(and other quantum crystals) under increasing pressure? As compression 
is raised the repulsive electrostatic interactions between neighboring 
electron clouds increase and consequently the atoms remain closer to their 
equilibrium positions in order to minimize their potential energy, $E_{\rm pot}$.  
By other side, due to the non-commutativity between the position and momentum 
quantum operators, whenever atomic localization increases so does the kinetic energy, 
$E_{\rm kin}$. 
Namely, pressure acts by incrementing both $E_{\rm pot}$ and $E_{\rm kin}$ energies 
and it is not explicitly known how the $|E_{\rm kin} / E_{\rm pot}|$ ratio, which 
can be regarded as a quantum level indicator of the system, evolves under compression. 
Answering to this and other similar questions is of paramount importance 
for modeling of materials in Earth and planetary sciences, since light 
weight species like $^{4}$He and H$_{2}$ are believed to be abundant in the interior 
of celestial bodies. More precisely, determining the exact role of quantum nuclear 
effects in compressed quantum crystals will permit to fully justify or disapprove 
the use of approximate approaches, routinely employed in high-pressure studies 
(e.g., Debye and quasi-harmonic 
models),~\cite{cazorla13,shevlin12,cazorla12,cazorla10,cazorla09b,freiman09,freiman12,freiman13} 
for estimation of ``zero-point energies'' and other related quantities
(e.g., phase transitions and atomic structure).
 
Quantifying the exact evolution of the energy in $^{4}$He under pressure, 
however, is not a simple task. Since compression profoundly affects the 
electronic structure of materials, the interaction models which at 
low pressures describe successfully $^{4}$He atoms or H$_{2}$ 
molecules turn out to be unreliable at high-$P$ conditions. 
This fact seriously hinders the application of quantum Monte Carlo 
methods to their study. From an ideal point of view, one would like to describe 
both the electronic and nuclear degrees of freedom in quantum crystals fully 
from first principles, that is, without relying on any substantial approximation 
to the atomic interactions. Nevertheless, such a strategy, although in principle 
is technically possible, it would require of an enormous amount of computational 
effort. Thus, in practice effective and simpler quantum simulation methods able 
to deal with large systems (i.e., composed of $100-1,000$ atoms) are highly desirable.

In this work, we present a comprehensive computational study of the energy and 
structural properties of hcp $^{4}$He at $T=0$~K and pressures up to 
$\sim 160$~GPa, based exclusively on first-principles methods. In particular, we employ 
density functional theory (DFT) to access the electronic band-structure of the crystal 
and the diffusion Monte Carlo (DMC) method to solve the time-dependent Schr\"odinger 
equation that renders the behavior of the quantum nuclei. The effective pair interaction 
between nuclei is constructed by fitting the static compression curve obtained with DFT 
to an analytical function based on the Aziz potential~\cite{aziz87} and an attenuation 
repulsion factor proposed by Moraldi.~\cite{moraldi12} We find that the 
$|E_{\rm kin} / E_{\rm pot}|$ ratio in solid $^{4}$He is overall depleted with increasing 
pressure due to a very small~(large) increase of $E_{\rm kin}$~($E_{\rm pot}$) at compressions 
larger than $\sim 85$~GPa. In particular, the $^{4}$He kinetic energy increases by no more 
than $\sim 15$~K in the pressure interval $85 \le P \le 150$~GPa.  
Such a small $E_{\rm kin}$ increase illustrates the unique ability of $^{4}$He atoms 
to remain extraordinarily delocalized within extremely dense environments as a result of
their quantum correlations. Accordingly, we find that the Lindenmann ratio in compressed 
$^{4}$He ($P \geqslant 15$~GPa) amount to $0.10$ almost independently of the pressure. 
Furthermore, we assess the performance of approximate quasi-harmonic DFT methods in 
evaluation of kinetic energies at $T = 0$~K and find that, in the 
best of the cases, these approaches exceedingly overestimate $E_{\rm kin}$ by 
$\sim 50$~\%. Quasi-harmonic approaches also turn out to be inadequate to 
describe the size of the $^{4}$He displacements around their equilibrium lattice 
positions. Thus, we resolve that quasi-harmonic DFT methods are not able to 
describe the ground-state properties of dense helium correctly.  
The main conclusions presented in this work can be extended to other light 
and weakly interacting species like, for instance, H$_{2}$, methane (CH$_{4}$) and 
ammonia (NH$_{3}$), wherein quantum nuclear effects are expected to be critically 
important.~\cite{geneste12,gao10,loubeyre96,mcmahon12,pickard08} 

The organization of this article is as follows. In the next section, we briefly 
explain the fundamentals of the methods employed and provide the technical details 
in our calculations. There, we present also our modeling strategy of the atomic 
interactions in solid $^{4}$He at high $P$. Next, we present our results for 
the equation of state, $|E_{\rm kin} / E_{\rm pot}|$ ratio, and structural properties of 
solid helium, together with some discussion. Finally, we summarize our main findings in 
Sec.~\ref{sec:conclusions}.

\section{Computational Methods}
\label{sec:methods}
In this work, density functional theory (DFT) provides the basis for our understanding of the 
electronic structure of solid $^{4}$He under pressure. In particular, 
we use the DFT output to construct an effective pairwise potential that makes it possible to 
simulate quantum helium crystals with the diffusion Monte Carlo (DMC) method
at low computational cost. In the next subsections, we briefly explain the basics of the DFT and 
DMC methods and present our proposed and easy-to-implement parametrization of the 
$^{4}$He--$^{4}$He interactions at high pressures (i.e., up to $\sim 160$~GPa). Also we 
review the main ideas of the quasi-harmonic approach, which is customarily employed  
for the estimation of zero-temperature kinetic energies in computational high-$P$ studies.

\subsection{Density Functional Theory}
\label{subsec:DFT}
DFT is a first-principles approach which has allowed for accurate and reliable 
knowledge of a great deal of materials with exceptional computational 
affordability.~\cite{martin04,kohanoff06} There is only one uncontrollable approximation 
in DFT, namely the functional used for the exchange-correlation energy $E_{\rm xc}$.
There is abundant evidence showing that commonly used $E_{\rm xc}$ functionals yield 
accurate results for a range of properties of metallic and non-metallic crystals, 
including the equilibrium lattice parameter, elastic constants, phonon frequencies, 
$T = 0$ equation of state (EOS) and solid-state phase boundaries.~\cite{cazorla08a,gillan06,cazorla08c}

It must be noted, however, that standard DFT methods do not describe properly 
long-range dispersive interactions in solids, like for instance van der Waals (vdW) forces, 
due to the local nature of the employed $E_{\rm xc}$ approximations.~\cite{dion04,basanta05}
Also, it is well-known that such a type of interactions plays a critical role in the 
cohesion of helium at low pressures. Nevertheless, short-range effects in rare-gas systems 
become increasingly more relevant as pressure is raised. Consequently, the description 
of helium and other similar materials attained with standard DFT becomes progressively 
more accurate as density is increased.~\cite{drummond06,nabi05}
In spite of this fact, we explicitly treat dispersive interactions in this work by 
employing Grimme's vdW approach~\cite{grimme06} and the  
exchange-correlation functional due to Perdew \textit{et al.}~\cite{perdew96} (hereafter 
denoted as PBE-vdW). As it will be shown later, considering long-range vdW interactions 
in our calculations has imperceptible effects on the final conclusions. 

A completely separate issue from the choice of $E_{\rm xc}$ is the implementation of 
DFT, which mainly concerns to the way in which electron orbitals are represented. 
Here, we have chosen the PAW ansatz~\cite{blochl94,kresse99} as implemented in the VASP 
code~\cite{kresse96} since it has been demonstrated to be greatly efficient.~\cite{cazorla07,taioli07}
Regarding other technical aspects in our DFT calculations, the electronic wave functions 
were represented in a plane-wave basis truncated at $500$~eV, and for integrations 
within the first Brillouin zone (BZ) we employed dense $\Gamma$-centered $k$-point 
grids of $14 \times 14 \times 14$. By using these parameters we obtained interaction 
energies that were converged to within $5$~K per atom. Geometry relaxations were 
performed by using a conjugate-gradient algorithm that kept the volume of the unit cell 
fixed while permitting variations of its shape, and the imposed tolerance on the atomic 
forces was $0.005$~eV$\cdot$\AA$^{-1}$. With this DFT setup, we calculated the total
energy of solid $^{4}$He in the volume interval $3 \le V \le 16$~\AA$^{3}$/atom.

\subsection{Zero-point energy within the quasi-harmonic approach}
\label{subsec:quasiharm}

\begin{figure}
\centerline
        {\includegraphics[width=1.0\linewidth]{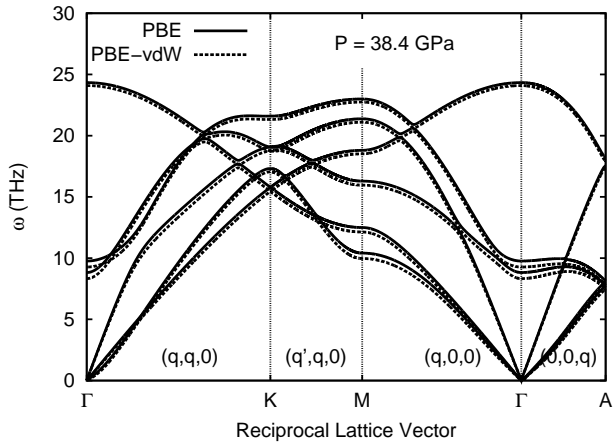}}
\caption{Phonon spectrum of solid $^{4}$He calculated at high pressure with DFT using  
         two different exchange-correlation energy functionals, one of which takes into
         account long-range attractive van der Waals interactions (see text).} 
\label{phonons}
\end{figure}

In the quasi-harmonic (QH) approach, one assumes that the potential energy
of a crystal can be approximated with a quadratic expansion around the
equilibrium atomic configuration of the form 
\begin{equation} 
E_{\rm harm} = E_{\rm eq} + \frac{1}{2}
\sum_{l\kappa\alpha,l'\kappa'\alpha'}
\Phi_{l\kappa\alpha,l'\kappa'\alpha'} u_{l\kappa\alpha}
u_{l'\kappa'\alpha'}~,
\label{eq:eqh}
\end{equation}
where $E_{\rm eq}$ is the total energy of the undistorted lattice,
$\boldsymbol{\Phi}$ the force-constant matrix, and $u_{l\kappa\alpha}$
is the displacement along Cartesian direction $\alpha$ of the atom
$\kappa$ at lattice site $l$. Usually, the associated dynamical problem 
is tackled by introducing
\begin{equation}
u_{l\kappa\alpha}(t) = \sum_{q} u_{q\kappa\alpha} \exp{ \left[ i
    \left(\omega t - \boldsymbol{q} \cdot (\boldsymbol{l}+
    \boldsymbol{\tau}_{\kappa} \right) \right] }~,
\end{equation}
where $\boldsymbol{q}$ is a wave vector in the first Brillouin zone (BZ)
defined by the equilibrium unit cell;
$\boldsymbol{l}+\boldsymbol{\tau}_{\kappa}$ is the vector that locates
the atom $\kappa$ at cell $l$ in the equilibrium structure. Then, the
normal modes are found by diagonalizing the dynamical matrix
\begin{equation}
\begin{split}
& D_{\boldsymbol{q};\kappa\alpha,\kappa'\alpha'} =\\ &
  \frac{1}{\sqrt{m_{\kappa}m_{\kappa'}}} \sum_{l'}
  \Phi_{0\kappa\alpha,l'\kappa'\alpha'} \exp{\left[
      i\boldsymbol{q}\cdot(\boldsymbol{\tau}_{\kappa}-\boldsymbol{l'}-\boldsymbol{\tau}_{\kappa'})
      \right]}~,
\end{split}
\end{equation}
and thus the material is treated as a collection of non-interacting
harmonic oscillators with frequencies $\omega_{\boldsymbol{q}s}$
(positively defined and non-zero) and energy levels
\begin{equation}
E^{n}_{\boldsymbol{q}s} = \left( \frac{1}{2} + n \right)
\omega_{\boldsymbol{q}s}~,
\end{equation}
where $0 \le n < \infty$. Within this approximation, the Helmholtz
free energy of a crystal with volume $V$ at temperature $T$ is given by
\begin{equation}
F_{\rm harm} (V,T) = \frac{1}{N_{q}}~k_{B} T \sum_{{\bf
    q}s}\ln\left[ 2\sinh \left( \frac{\hbar\omega_{{\bf
        q}s}(V)}{2k_{\rm B}T} \right) \right]~,
\label{eq:fharm}
\end{equation}
where $N_{q}$ is the total number of wave vectors used in the 
BZ integration, and the explicit $V$-dependence of the frequencies
is indicated.
In the limit of zero-temperature Eq.~(\ref{eq:fharm}) transforms into
\begin{equation}
E_{\rm harm} (V) = \frac{1}{N_{\rm q}} \sum_{\boldsymbol{q}s}
\frac{1}{2}\hbar\omega_{\boldsymbol{q}s}(V)~,
\label{eq:zpe}
\end{equation}
which usually is referred to as the ``zero-point energy'' (ZPE).
We note that in many computational high-$P$ studies ZPE corrections
turn out to be decisive in the prediction of accurate transition pressures 
which involve two crystal structures with similar $E_{\rm eq}$ 
energies.~\cite{cazorla13,shevlin12,cazorla08}

In order to compute the QH free energy of a crystal, it is necessary
to know its full phonon spectrum. For this, we employ here the ``direct approach'' 
and DFT calculations. In the direct approach 
the force-constant matrix is directly calculated in real-space by considering the 
proportionality between the atomic displacements and forces when the former are 
sufficiently small.~\cite{kresse95,alfe01} In this case, large supercells have to 
be constructed in order to guarantee that the elements of the force-constant 
matrix have all fallen off to negligible values at their boundaries, a 
condition that follows from the use of periodic boundary conditions.~\cite{alfe09} 
Once the force-constant matrix is obtained, we can Fourier-transform it 
to obtain the phonon spectrum at any $q$-point.

The quantities with respect to which our QH DFT calculations need to be 
converged are the size of the supercell, the size of the atomic displacements, 
and the numerical accuracy in the calculation of the atomic forces and BZ sampling. 
We found the following settings to fulfill convergence of ZPE corrections to 
within $5$~K/atom: $4 \times 4 \times 3$ supercells (i.e., $48$ repetitions of 
the hcp unit cell containing a total of $96$ atoms), atomic displacements of $0.02$~\AA, 
and special Monkhorst-Pack~\cite{monkhorst76} grids of $12 \times 12 \times 12$ $q$-points 
to compute the sums in Eq.~(\ref{eq:fharm}). 
Regarding the calculation of the atomic forces with VASP, we found that the 
density of $k$-points had to be increased slightly with
respect to the value used in the energy calculations (i.e., from $14
\times 14 \times 14$ to $16 \times 16 \times 16$) and that computation of the 
non-local parts of the pseudopotential contributions had to be performed in
reciprocal, rather than real, space. These technicalities were adopted
in all our force-constant matrix calculations.
The value of the phonon frequencies and ZPE energies were obtained with the 
PHON code due to Alf\`e.\cite{alfe09} In using this code, we exploited the 
translational invariance of the system to impose the three acoustic branches to 
be exactly zero at the $\Gamma$ $q$-point, and used central differences in the 
atomic forces (i.e., we considered positive and negative atomic displacements). 
As an example of our phonon frequency calculations, we show in Fig.~\ref{phonons} 
the full $^{4}$He phonon spectrum computed at $P \sim 40$~GPa. It is noted
that the effect of considering van der Waals forces there is remarkably small.

\subsection{Construction of the effective interatomic potential $V_{\rm Aziz-B}$}
\label{subsec:fitpot}

\begin{figure}
\centerline
        {\includegraphics[width=1.0\linewidth]{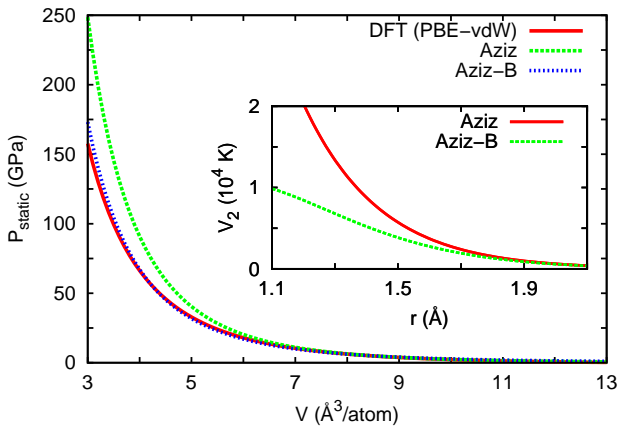}}
\caption{Calculated static equation of state of solid helium (i.e., considering  
         immobile nuclei at the equilibrium lattice positions) using DFT (PBE-vdW) and two 
         effective pairwise interaction models, namely the well-known Aziz potential and 
         the modified version Aziz-B. \emph{Inset}: Comparison of the repulsive cores of 
         the Aziz and Aziz-B potential models.} 
\label{fitpot}
\end{figure}

In a previous work, we demonstrated that the semi-empirical pairwise potential due to 
Aziz~\cite{aziz87} is inadequate to describe solid $^{4}$He at pressures higher than 
$\sim 1$~GPa.~\cite{cazorla08a} The Aziz potential, $V_{\rm Aziz} (r)$ (where $r$ represents 
the radial distance between a pair of atoms), is composed of two basic contributions: 
(1)~$V_{\rm rep} (r)$ which is short-ranged and repulsive and accounts for the electrostatic 
and Pauli-like interactions between close electrons, and (2)~$V_{\rm bond} (r)$ that is 
long-ranged and attractive and describes the interactions between instantaneous and 
induced multipoles created in the electron clouds. As pressure is raised electronic repulsion 
prevails over attraction, namely $V_{\rm rep} \gg V_{\rm bond}$. In the Aziz case, 
however, we found that $V_{\rm rep}$ is  unrealistically too large at small $r$. In order 
to ammend this flaw, we performed a series of DFT energy calculations considering 
different configurations in which $^4$He atoms are fixed on their equilibrium hcp 
positions. Subsequently, we fitted a modified version of the Aziz potential, hereafter 
denoted as $V_{\rm Aziz-B} (r)$, to our DFT results. The form of this modified Aziz potential 
is based on the model proposed by Moraldi for solid H$_{2}$,~\cite{moraldi12} which reads
\begin{equation}
V_{\rm Aziz-B} (r) = V_{\rm rep} (r) \cdot f_{\rm att} (r) + V_{\rm bond} (r)~.
\label{aziz-b}
\end{equation}      
In the above equation, $V_{\rm rep}$ and $V_{\rm bond}$ are the original 
repulsive and attractive parts found in the Aziz potential, and $f_{\rm att}$ is an 
attenuation repulsion factor of the form
\begin{eqnarray}
&& f_{\rm att} (r) =  \exp{\left[-A_{\rm att} \left(\frac{R_{\rm att}}{r} -1 \right)^{C_{\rm att}}\right]} ~ r \le R_{\rm att}   \nonumber  \\ 
&& \qquad  \qquad \qquad \qquad \qquad   1   \qquad \qquad \qquad ~ r > R_{\rm att}     
\label{eq:attenuation}
\end{eqnarray} 
where $A_{\rm att}$, $R_{\rm att}$, and $C_{\rm att}$ are parameters to be determined.

In our fitting strategy, rather than trying to match the set of calculated DFT 
energies, we pursued to reproduce the static DFT equation of state (i.e., 
$P_{\rm static} = -d E_{\rm DFT} (V) /d V$). In fact, the physics contained in any 
pair of potential functions $V_{2} (r)$ and $V_{2}'(r) = V_{2}(r) + V_{0}$ (where 
$V_{0}$ is a constant) is the same, hence the truly important quantities to reproduce 
are variations of the total energy with respect to the positions of the atoms 
(e.g., pressure and atomic forces). For this, 
we fitted the DFT energies to a third order Birch-Murnaghan equation 
of the form~\cite{birch78}
\begin{eqnarray} 
&&E_{\rm eq} (V) = E_{0} + \frac{3}{2}~V_{0}~B_{0} \times \nonumber \\  
&& \bigg [ -\frac{\chi}{2} \left ( \frac{V_{0}}{V} \right )^2 + \frac{3}{4}~ \left ( 1+2 \chi \right ) \left ( \frac{V_{0}}{V} \right )^{(4/3)} \nonumber \\
&& - \frac{3}{2} \left ( 1+\chi \right ) \left (\frac{V_{0}}{V} \right )^{(2/3)} + \frac{1}{2} \left (\chi+\frac{3}{2}\right ) \bigg ]
\label{eq:eqstate}
\end{eqnarray}
[where $E_{0}$ and $B_{0}= V_{0}\frac{d^2E}{dV^2}$ are the values of the
energy and bulk modulus at the equilibrium volume $V_{0}$, respectively,
$\chi = \frac{3}{4}\left ( 4 - B^{'}_{0} \right )$ and ~$B^{'}_{0}=\left(
d B_{0}/d P\right)$, with derivatives evaluated at zero pressure] and 
then searched iteratively for the $f_{\rm att}$ parameters which better 
reproduced the DFT $P_{\rm static} (V)$ curve. 
In Fig.~\ref{fitpot} we show our best fit results, which correspond to 
values $A_{\rm att} = 0.95$,  $R_{\rm att} = 2.34$~\AA~, and $C_{\rm att} = 1.50$.
This constitutes our choice for the effective Aziz-B potential in the remainder 
of this article. 
In the same figure we also compare the repulsive core of the original and 
modified Aziz potentials, where the corresponding attenuation effect is clearly
appreciated. We note that due to the specific form of $f_{\rm att}$, 
the $V_{\rm Aziz-B} (r)$ potential displays a positive slope 
at radial distances $0 \le r \le d \sim 1.0$~\AA~. This feature is manifestly
incorrect from a physical point of view.~\cite{boninsegni13} 
Nevertheless, at the highest pressure considered in this work (i.e., $\sim 160$~GPa) 
the $^{4}$He atoms remain separated by distances of about $1.6$~\AA~, hence 
we are safely distant from exploring the unphysical region $r \le d$ in our
simulations (as we have indeed checked; see also Sec.~\ref{sec:results}).

\subsection{Diffusion Monte Carlo}
\label{subsec:DMC}

\begin{figure}
\centerline
        {\includegraphics[width=1.0\linewidth]{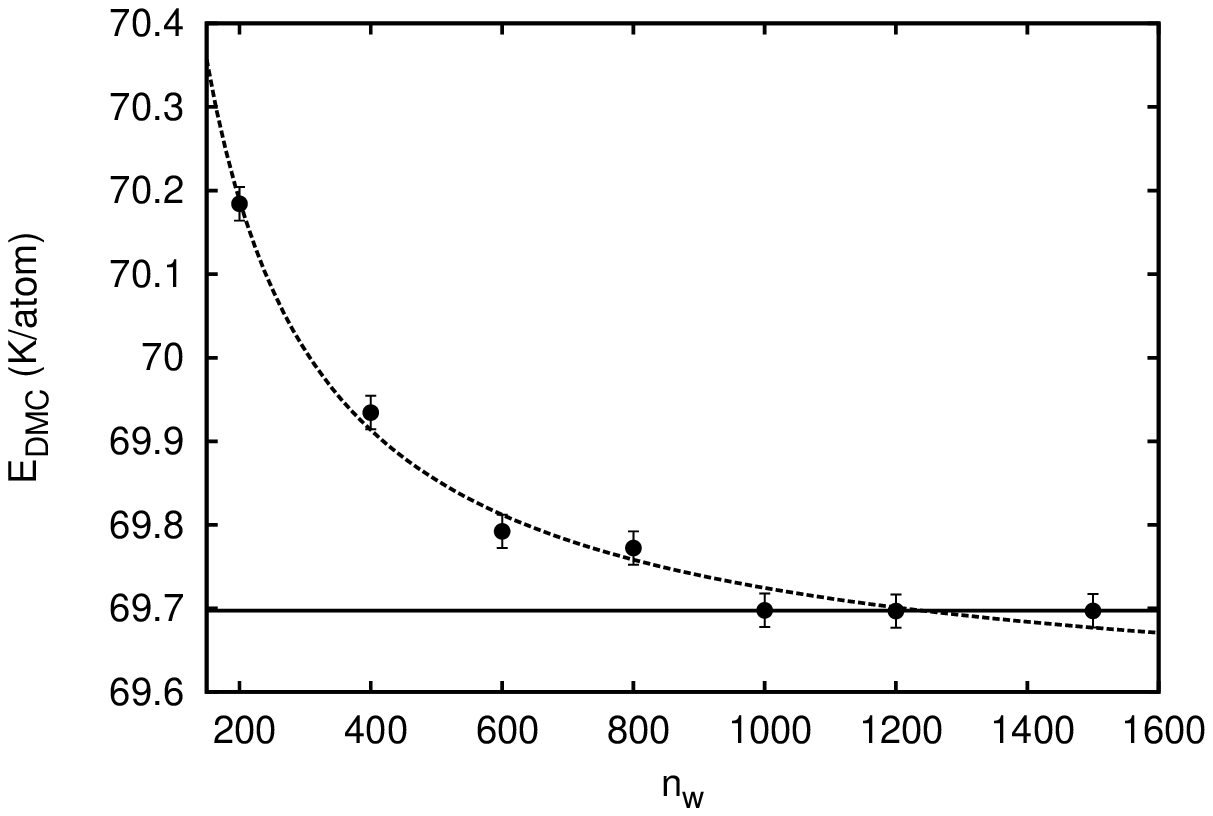}}
        {\includegraphics[width=1.0\linewidth]{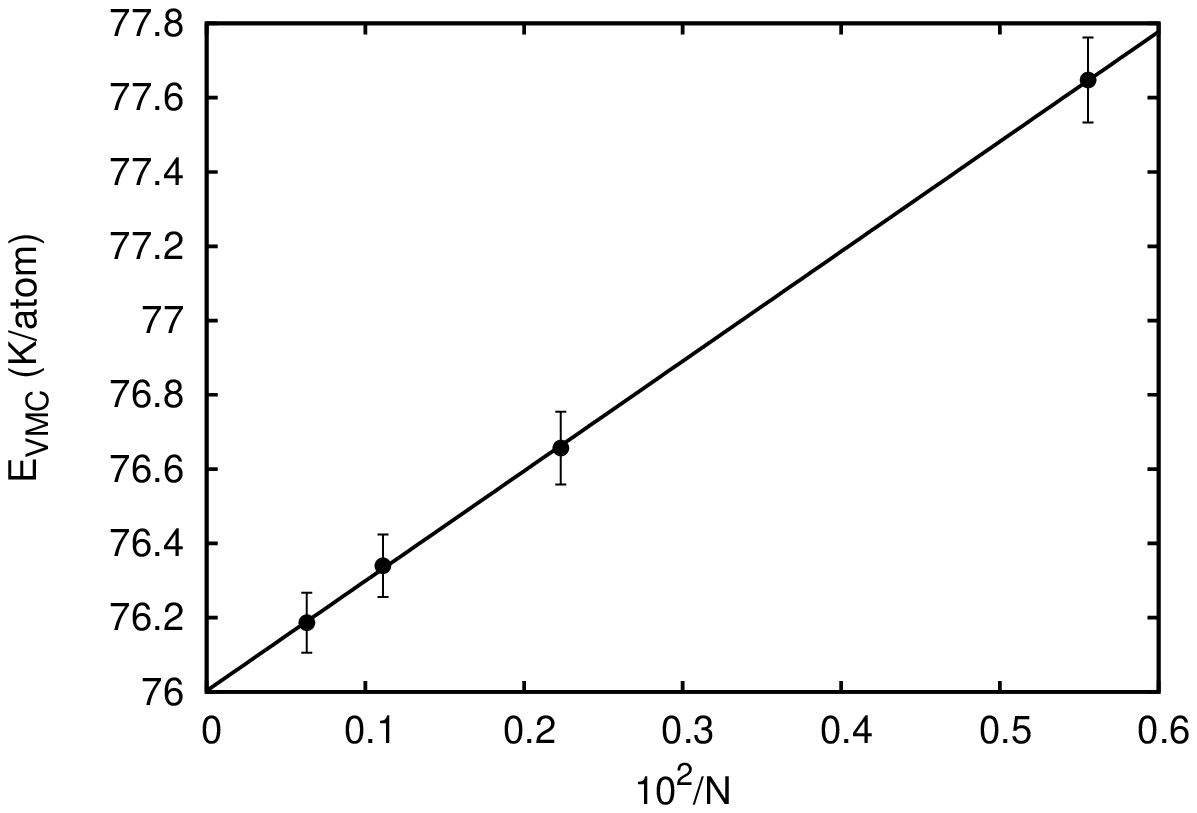}}
\caption{(a)~Dependence of the ground-state energy calculated with the DMC method  
          on the critical population of walkers. The dashed line corresponds to 
          a (monotonically decreasing) inverse power-law fitted to the calculated 
          $E_{\rm DMC} (n_{w})$ energies, whereas the horizontal solid line marks 
          the plateau that is reached at $1000 \le n_{w}$ values. (b)~Dependence of the
          total energy calculated with the VMC method on the number of particles. 
          Extrapolation to the thermodynamic limit is achieved through a linear fit.}
\label{dmcbias}
\end{figure}

DMC is an accurate computational method that provides the \emph{exact} 
(within statistical errors) ground-state energy of a bosonic many-body 
interacting system.~\cite{hammond94,guardiola98,ceperley79} 
This technique is based on a short-time approximation for the Green's function, 
corresponding to the imaginary time-dependent Schr\"{o}dinger equation, which is 
solved up to a certain order of accuracy within an infinitesimal interval $\Delta \tau$. 
Despite this method is algorithmically simpler than domain Green's function Monte 
Carlo,~\cite{ceperley79,kalos74} it presents some $\left(\Delta \tau\right)^{n}$
bias coming from the factorization of the imaginary time propagator 
$e^{-\frac{\Delta\tau}{\hbar}{\rm H}}$.
Our DMC implementation is quadratic,~\cite{chin90} hence the time-step 
bias is efficiently controlled by choosing a sufficiently small $\Delta \tau$.

The Hamiltonian, ${\rm H}$, describing our system is
\begin{equation}
\label{eq:hamiltonian}
{\rm H} = - \frac{\hbar^{2}}{2m_{\rm He}} \sum_{i=1}^{N} \nabla^{2}_{i} +
       \sum_{i<j}^{N} V_{2}^{\rm eff}(r_{ij})~,
\end{equation}
where $m_{\rm He}$ is the mass of a $^4$He atom, $r_{ij}$ the distance between atoms 
composing a $i$,$j$ pair, and $V_{2}^{\rm eff}(r_{ij})$ a pairwise
interatomic model (i.e., $V_{\rm Aziz}$ and $V_{\rm Aziz-B}$, see Sec.~\ref{subsec:fitpot}).
The corresponding Schr\"odinger equation in imaginary time ($it \equiv \tau$) is 
\begin{equation}
\label{eq:schrodinger}
-\hbar\frac{\partial \Psi({\bf r},\tau)}{\partial \tau}= \left({\rm H}-E\right)\Psi({\bf r},\tau)~, 
\end{equation}
where $E$ is an arbitrary constant. Eq.~(\ref{eq:schrodinger}) can be formally 
solved by expanding the solution $\Psi({\bf r}, \tau)$ in the basis set of the energy 
eigenfunctions $\{\phi_{n}\}$. At large imaginary time $\Psi({\bf r}, \tau)$ tends 
to the ground state wave function $\phi_{0}$ and the expected value of the Hamiltonian 
to the ground-state energy $E_{0}$. The hermiticity of the Hamiltonian guarantees the 
equality   
\begin{equation}
\label{eq:groundstate1}
E_{0} = \frac{\left<\phi_{0}|{\rm H}|\phi_{0}\right>}{\left<\phi_{0}|\phi_{0}\right>}=
\frac{\left<\phi_{0}|{\rm H}|\psi_{T}\right>}{\left<\phi_{0}|\psi_{T}\right>}~, 
\end{equation}
where $\psi_{T}$ is a convenient guiding wave function that depends on the atomic 
coordinates of the system ${\bf r}\equiv \{ {\bf r}_{1}, {\bf r}_{2},...,{\bf r}_{N} \}$.
Consequently, the ground-state energy is obtained in practice by computing with 
stochastic techniques the integral
\begin{equation}
\label{eq:integral}
E_{\rm DMC}= \lim_{\tau \to\infty} \frac{1}{\cal{N}} \int_{V} E_{L}\left({\bf 
r}\right) f\left({\bf r},\tau\right) d{\bf r} = E_{0}~,
\end{equation} 
where $f\left({\bf r},\tau\right)=\Psi\left({\bf 
r},\tau\right)\psi_{T}\left({\bf r}\right)$, $\cal{N}$ is a normalization 
factor, and $E_{L}\left({\bf r}\right)$ is the local energy defined as
${\rm H}\psi_{T}\left({\bf r}\right)/\psi_{T}\left({\bf r}\right)$. 
The introduction of the guiding wave function $\psi_{T}\left({\bf 
r}\right)$ in $f\left({\bf r},\tau\right)$, known as importance sampling, 
reduces significantly the variance of the integral (\ref{eq:integral}) 
[for instance, by imposing $\psi_{T}\left({\bf r}\right)=0$ 
when $r_{ij}$ is very small].

In this work, the guiding wave function used for importance sampling corresponds 
to the extensively tested Nosanow-Jastrow model~\cite{nosanow64,hansen68,hansen69}
\begin{equation}
\psi_{\rm NJ}\left({\bf r}_{1},{\bf r}_{2},...,{\bf r}_{N}\right) =
\prod_{i\neq j}^{N} {\rm f_{2}}(r_{ij}) \prod_{i=1}^{N}{\rm g_{1}}(|{\bf r}_{i}-
{\bf R}_{i}|)~,
\label{eq:nosanojashe}
\end{equation}
with ${\rm f_{2}}(r) = e^{-\frac{1}{2}\left(\frac{b}{r}\right)^{5}}$ and 
${\rm g_{1}}(r) = e^{-\frac{1}{2}ar^{2}}$, and where $a$ and $b$ are variational 
parameters. This model is composed of two-body correlation functions ${\rm f_{2}}(r)$ 
deriving from the interatomic potential, and one-body functions ${\rm g_{1}}(r)$ 
that localize the particles around the positions of the equilibrium  
lattice $\lbrace {\bf R}_{i}\rbrace$. The Nosanow-Jastrow model is not 
Bose symmetric under the exchange of particles however $\psi_{\rm NJ}$ has been shown 
to provide very accurate energy and structure results in DMC simulations.~\cite{cazorla09} 
We note that the parameters contained in $\psi_{\rm NJ}$ are optimized with the 
variational Monte Carlo technique (VMC) at each considered density.~\cite{hammond94}
For instance, at $\rho = 0.06$~\AA$^{-3}$~ we obtain $b = 2.94$~\AA~ and $a = 3.21$~\AA$^{-2}$~, 
and at $\rho = 0.33$~\AA$^{-3}$~, $b = 1.84$~\AA~ and $a = 29.08$~\AA$^{-2}$~. 

Our DMC calculations need to be converged with respect to the time step
$\Delta \tau$, critical population of walkers $n_{w}$, and number of particles
$N$. We have adjusted $\Delta \tau$ and $n_{w}$ in order to eliminate any 
possible bias coming from them. In particular, these are $10^{-4}$~K$^{-1}$ 
and $10^3$, respectively. In Fig.~\ref{dmcbias}a, we demonstrate that the 
selected $n_{w}$ value perfectly guarantees proper convergence of the total 
ground-state energy. In fact, we do not observe the monotonically decreasing
$1/r$ law reported in Ref.~[\onlinecite{boninsegni12}] (see Fig.~\ref{dmcbias}a).
Finite size errors have been corrected by following the variational approach 
introduced in Ref.~[\onlinecite{cazorla08a}], which proved to be very accurate 
in describing solid $^4$He at moderate pressures. Namely, the total ground-state 
energy of the system is computed as $E_{\rm DMC}(\infty) = 
E_{\rm DMC}(N_{0}) + \Delta E_{\rm VMC} ^{\rm tail} (N_{0})$, where 
\begin{equation}
\label{eq:vtc}
\Delta E_{\rm VMC} ^{\rm tail} (N_{0}) = E_{\rm VMC}^{\infty} - E_{\rm VMC}^{N_{0}}~.  
\end{equation}
In the equation above energy superscripts indicate number of particles, 
$N_{0} = 180$ is the number of atoms employed in the DMC simulations, and 
$E_{\rm VMC} \equiv \left< \psi_{\rm NJ}|{\rm H}|\psi_{\rm NJ} \right> / 
\left< \psi_{\rm NJ}|\psi_{\rm NJ}\right>$ is the variational energy 
calculated with the guiding wave function~(\ref{eq:nosanojashe}).
The variational energy in the $N \to \infty$ limit, $E_{\rm VMC}^{\infty}$, is estimated 
by successively enlarging the simulation box (i.e., up to $1584$ particles) at fixed
density and performing a linear extrapolation to infinite volume. 
Indeed, this procedure turns out to be computationally affordable within VMC but not  
within DMC. In Fig.~\ref{dmcbias}b, we show a test case in which the
adequacy of the $E_{\rm VMC} (N)$ linear extrapolation is shown.

\section{Results and Discussion}
\label{sec:results}

\begin{figure}
\centerline
        {\includegraphics[width=1.0\linewidth]{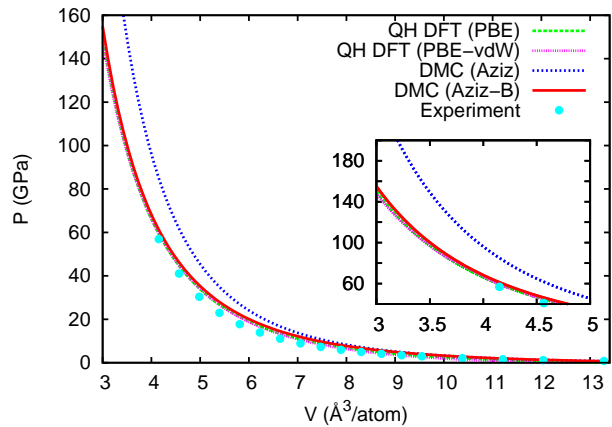}}
\caption{Zero-temperature equation of state of helium calculated with different methods  
         and considering exact (i.e., Aziz and Aziz-B) and approximate (i.e., PBE and 
         PBE-vdW) estimation of quantum nuclear effects. 
         Experimental data found in Ref.~[\onlinecite{loubeyre93}] 
         are shown for comparison. \emph{Inset}: The high-$P$ region in the EOS is 
         zoomed in to appreciate better the differences.}
\label{eos}
\end{figure}

\begin{figure}
\centerline
        {\includegraphics[width=1.0\linewidth]{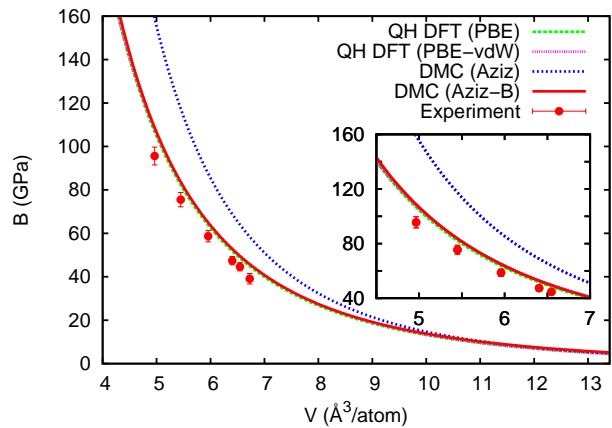}}
\caption{Calculated $^{4}$He bulk modulus with different methods and (i.e., 
         Aziz and Aziz-B) and considering approximate (i.e., PBE and PBE-vdW)  
         estimation of quantum nuclear effects. Experimental data found in 
         Ref.~[\onlinecite{zha04}] are shown for comparison. 
         \emph{Inset}: The high-$P$ $B$ region is zoomed in to appreciate 
         better the differences.}
\label{bulk-modulus}
\end{figure}

In Fig.~\ref{eos}, we show the calculated equation of state (EOS) in solid 
$^4$He using (i)~DFT and quasi-harmonic zero-point energy corrections
[i.e., curves PBE and PBE-vdW where $E_{\rm DFT} (V) = E_{\rm eq} (V) + E_{\rm harm} (V)$, 
see Eqs.~(\ref{eq:zpe}) and~(\ref{eq:eqstate})], and (ii)~DMC with the  
effective pairwise potentials $V_{\rm Aziz}$  and $V_{\rm Aziz-B}$. For comparison 
purposes, we include also experimental data from Ref.~[\onlinecite{loubeyre93}].  
Very good agreement is found between experiments and the calculated DMC(Aziz-B) and 
QH DFT equations of state. In contrast, results obtained with the original Aziz potential 
and the DMC method largely overestimate the measured pressures (as it was already expected,
see Sec.~\ref{subsec:fitpot}).
The $P(V)$ curves shown in Fig.~\ref{eos} are based on the $E(V)$ parametrization
introduced in Eq.~(\ref{eq:eqstate}). The optimal parameters obtained in the 
DMC(Aziz-B) case are: $V_{0}^{\rm DMC} = 15.92$~\AA$^{3}$, $B_{0}^{\rm DMC} = 2.66$~GPa, 
and $\chi^{\rm DMC} = -0.086$; and in the DFT(PBE-vdW) case: $V_{0}^{\rm DFT} = 
12.23$~\AA$^{3}$, $B_{0}^{\rm DFT} = 6.38$~GPa, and $\chi^{\rm DFT} = 0.026$~(relative 
errors associated to these quantities typically are $1-5$~\%). Upon inspection of  
Fig.~\ref{eos} one may arrive at the following conclusions: (i)~long-range van der Waals 
interactions are second order in compressed solid $^{4}$He thus there is not a 
real need to consider them in practical simulations, and (ii)~quasi-harmonic approaches 
based on DFT appear to be reliable methods for predicting zero-temperature EOS in 
compressed quantum crystals. 

In Fig.~\ref{bulk-modulus}, we enclose the bulk modulus of $^{4}$He, $B (V) = - V 
\left( d P / dV \right)_{V}$, calculated with the QH DFT and DMC methods (see also Table~I). 
Experimental data from Ref.~[\onlinecite{zha04}] are also shown for comparison. As in the 
previous case, we find notable agreement between the DMC(Aziz-B), QH DFT and experimental 
results, which further demonstrates the reliability of our devised pairwise 
potential model. We must note here that analysis of the elastic properties in dense helium 
is beyond the scope of the present work. In fact, it has been known for some time that  
in order to attain a realistic description of elasticity in rare gases under pressure, 
it is necessary to consider many-body interactions beyond pairwise.~\cite{pechenik08,grimsditch86} 
We therefore leave the study of these important physical quantities to future work.   

\begin{figure}
\centerline
        {\includegraphics[width=1.0\linewidth]{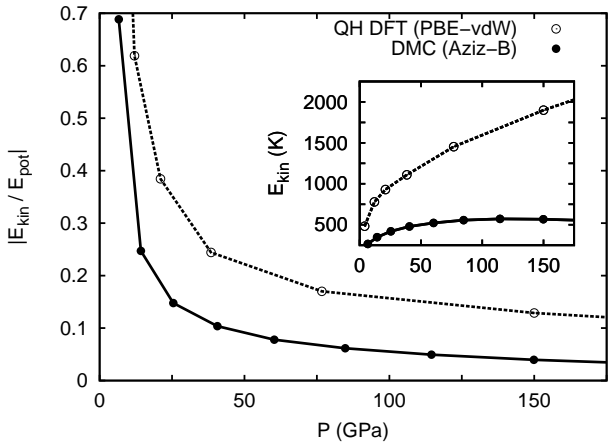}}
\caption{$|E_{\rm kin} / E_{\rm pot}|$ quantum indicator calculated considering exact 
         (DMC) and approximate (QH DFT) estimation of zero-temperature quantum
         nuclear effects, expressed as a function of pressure. \emph{Inset}:
         Zero-temperature kinetic energy of solid helium calculated with the
         DMC and quasi-harmonic DFT approaches, expressed as a function of
         pressure. The dashed and solid lines are guides-to-the-eye.}
\label{tvratio}
\end{figure}

\begin{figure}
\centerline
        {\includegraphics[width=1.0\linewidth]{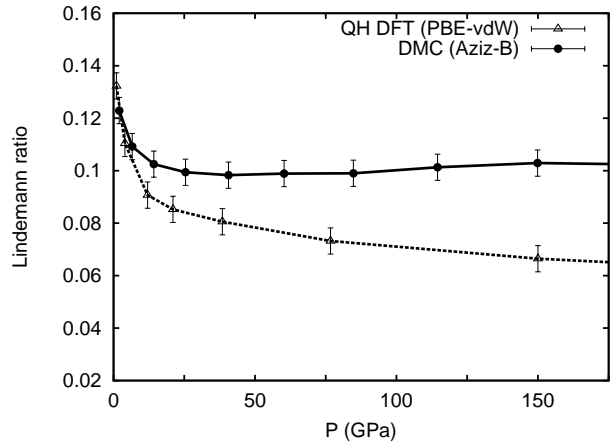}}
\caption{$^{4}$He Lindemann ratio calculated with exact (DMC) and approximate 
         (quasi-harmonic DFT) methods, expressed as a function of pressure. 
         The dashed and solid lines are guides-to-the-eye.}
\label{lindemann}
\end{figure}
    
As we mentioned in the Introduction, the $|E_{\rm kin} / E_{\rm pot}|$ ratio 
can be regarded as a qualitative indicator of the degree of quantumness of a condensed 
matter system at $T = 0$~K. Actually, the larger the kinetic energy the more important 
quantum nuclear effects are. For instance, in liquid $^{4}$He at the equilibrium density 
$E_{\rm kin}$ amounts to $14.6$~K, which is equal to the $\sim 67$~\% of 
the potential energy (in absolute value).~\cite{boronat94}
In the quasi-harmonic DFT approach, the $|E_{\rm kin} / E_{\rm pot}|$ ratio can be 
estimated as $|E_{\rm harm} / E_{\rm eq}|$ [see Eq.~(\ref{eq:zpe})]. Meanwhile, in
the DMC approach both $E_{\rm pot}$ and $E_{\rm kin} = E - E_{\rm pot}$ energies can 
be computed exactly (we note that for evaluation of $E_{\rm pot}$ we have employed the pure 
estimator technique~\cite{barnett91,casulleras95}) and hence so the $|E_{\rm kin} / E_{\rm pot}|$ 
ratio. In Fig.~\ref{tvratio}, we enclose our $|E_{\rm kin} / E_{\rm pot} |$ 
results obtained with the QH DFT and DMC methods and expressed as a function of
pressure (see also Table~I). 
There, it is shown that at pressures below $\sim 20$~GPa $^{4}$He behaves as an extreme 
quantum crystal, wherein the atomic kinetic energy is of the same order of magnitude
than the cohesive energy. We also find that the quantum character of solid helium, 
as quantified with the $|E_{\rm kin} / E_{\rm pot}|$ ratio, is progressively depleted 
with raising pressure. This occurs because the 
increase in potential energy caused by compression largely surpasses the accompanying 
increase in the kinetic energy (see Table~I). Actually, in Fig.~\ref{tvratio} we report
the explicit variation of the kinetic energy with pressure: it is found that $E_{\rm kin}$ 
increases noticeably from equilibrium up to compressions of $\sim 85$~GPa, however, at 
larger $P$ it just grows slightly (see Table~I). In particular, $E_{\rm kin}$ increases
by no more than $\sim 15$~K in the pressure interval $85 \le P \le 150$~GPa.
Such a tiny $P$-induced kinetic energy gain constitutes an original finding, and we will 
comment again on it in the next paragraphs. 
Meanwhile, we find that at any conditions the $|E_{\rm kin} / E_{\rm pot}|$ 
ratio calculated with QH DFT is significantly larger than the values obtained with DMC. 
In particular, the $E_{\rm harm} (P)$ curve displays always a large positive variation 
with increasing $P$ and it lies widely above $E_{\rm kin} (P)$. In fact, kinetic
energy discrepancies with respect to the DMC(Aziz-B) results amount to at least $\sim 50$~\% 
(see Fig.\ref{tvratio} and Table~I). 
These huge differences indicate that, despite QH DFT approaches may provide reasonable 
EOS (essentially because at high pressures $E_{\rm kin}$ is always small as compared 
to $E_{\rm pot}$), these cannot reproduce accurately quantum nuclear effects in dense helium.
This conclusion is of fundamental relevance to computational work done in high pressure
science, where ``zero-point energy'' corrections usually turn out to be decisive
in the prediction of phase transitions. Namely, according to our analysis QH DFT 
approaches may fail significantly at determining the $T = 0$~K phase diagram of substances
in which quantum nuclear effects are predominant.~\cite{geneste12,biermann98,kitamura00}

\begin{table*}
\begin{center}
\begin{tabular}{ c c c c c c c c }
\hline
\hline
 $ $ & $ $ & $ $ & $ $ & $ $ & $ $ & $ $ & $ $ \\
$\quad V$~(\AA$^{3}$) \quad & \quad $P$~(GPa) \quad &  \quad $B$~(GPa) \quad &  
\quad $E_{\rm DMC}$ \quad & \quad $\Delta E_{\rm VMC}^{\rm tail} 
(N_{0})$ \quad &  \quad $E_{\rm kin}$ \quad & \quad $E_{\rm harm}$ \quad 
 & \quad $\gamma$ \quad \\
 $ $ & $ $ & $ $ & $ $ & $ $ & $ $ & $ $ & $ $ \\
\hline
 $ $ & $ $ & $ $ & $ $ & $ $ & $ $ & $ $ & $ $ \\
 $11.13$& $2.0~(1)$ & $9.8~(1)   $  & $401.8~(1) $ & $-4.2~(1) $ & $174.6~(1)$ & $373~(5)$  & $ 0.12~(1)$ \\
 $8.35$ & $6.6~(1)$ & $24.1~(1)  $  & $1124.8~(1)$ & $-7.9~(1) $ & $265.9~(1)$ & $585~(5)$  & $ 0.11~(1)$ \\
 $6.68$ & $14.2~(1)$& $46.7~(1)  $  & $2281.1~(1)$ & $-15.8~(1)$ & $347.9~(1)$ & $831~(5)$  & $ 0.10~(1)$ \\
 $5.57$ & $25.4~(1)$& $78.6~(1)  $  & $3843.9~(1)$ & $-24.2~(1)$ & $419.9~(1)$ & $978~(5)$  & $ 0.10~(1)$ \\
 $4.77$ & $40.6~(1)$& $121.6~(1) $  & $5761.7~(1)$ & $-34.3~(1)$ & $479.0~(1)$ & $1132~(5)$ & $ 0.10~(1)$ \\
 $4.17$ & $60.3~(1)$& $176.6~(1) $  & $7971.7~(1)$ & $-47.5~(1)$ & $522.1~(1)$ & $1314~(5)$ & $ 0.10~(1)$ \\
 $3.71$ & $84.8~(1)$& $243.6~(1) $  & $10413.5~(1)$& $-64.2~(1)$ & $556.2~(1)$ & $1515~(5)$ & $ 0.10~(1)$ \\
 $3.34$ & $114.5~(1)$& $324.9~(1)$ & $13026.6~(1)$& $-81.7~(1)$ & $567.0~(1)$  &  $1714~(5)$ & $ 0.10~(1)$ \\
 $3.04$ & $149.9~(1)$& $419.9~(1)$ & $15758.4~(1)$& $-99.1~(1)$ & $571.5~(1)$  &  $1906~(5)$ & $ 0.10~(1)$ \\
 $ $ & $ $ & $ $ & $ $ & $ $ & $ $ & $ $ & $ $ \\
\hline
\hline
\end{tabular}
\end{center}
\caption{Energetic and structural properties of dense $^{4}$He computed with the DMC method 
         and Aziz-B pairwise model interaction (see text). $N_{0} = 180$ and stands for the 
         number of atoms employed in the DMC simulations. Zero-point energies, $E_{\rm harm}$, 
         obtained within the QH DFT approach are enclosed for comparison.   
         Energies are expressed in units of Kelvin and the statistical uncertainties 
         are within parentheses.} 
\end{table*}

The almost flat $E_{\rm kin} (P)$ curve obtained at $P \ge 85$~GPa constitutes an 
original, and to some extent unexpected, finding. Aimed at better understanding the origins 
of this effect, we computed the $^{4}$He Lindemann ratio $\gamma = \sqrt{\langle
{\bf u}^{2} \rangle}/a$ (where the quantity in the numerator represents the averaged 
mean squared displacement of the atoms taken with respect to their equilibrium hcp 
positions) as a function of pressure with the DMC and pure estimator 
techniques.~\cite{barnett91,casulleras95} 
The Lindemann ratio results enclosed in Fig.~\ref{lindemann} and Table~I show that 
the size of the $^{4}$He displacements around their equilibrium positions does not 
shrink appreciably with compression: $\gamma$ remains almost constant around $0.10$ 
at $P \geqslant 15$~GPa. This finding is consistent with the already disclosed 
$E_{\rm kin} (P)$ curve: $^{4}$He atoms can persist chiefly delocalized 
over ample pressure intervals in which their kinetic energy does not increase 
appreciably as a result of their quantum correlations. 
Meanwhile, $^{4}$He Lindenmann ratios calculated with the QH DFT approach 
(where $\langle {\bf u}_{\rm DFT}^{2} \rangle$ is estimated as $9 \hbar^{2} / 
8 m_{\rm He} E_{\rm harm}$~\cite{cazorla08d,arms03}) exhibit a monotonous decrease 
with increasing pressure and do not agree with the DMC results obtained at high $P$ 
(see Fig.~\ref{lindemann}). These large discrepancies show that structural details 
in dense quantum solids can neither be described correctly with QH DFT approches. 

\begin{figure}
\centerline
        {\includegraphics[width=1.0\linewidth]{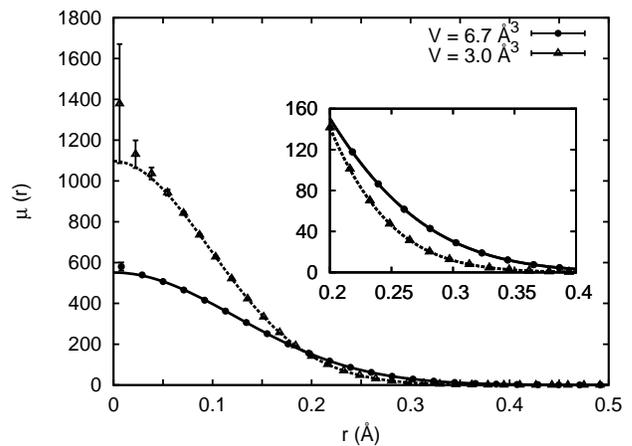}}
\caption{Radial atomic distribution function calculated with the DMC method and 
         pure estimator technique at $V = 6.7$~\AA$^{3}$ and $3.0$~\AA$^{3}$, 
         employing the modified Aziz-B pairwise potential. \emph{Inset}: 
         The intermediate region is zoomed in for a better appreciation of 
         the results. Lines correspond to Gaussian fits performed in the displayed
         $r$ interval.}
\label{profile}
\end{figure}

\begin{figure}
\centerline
        {\includegraphics[width=1.0\linewidth]{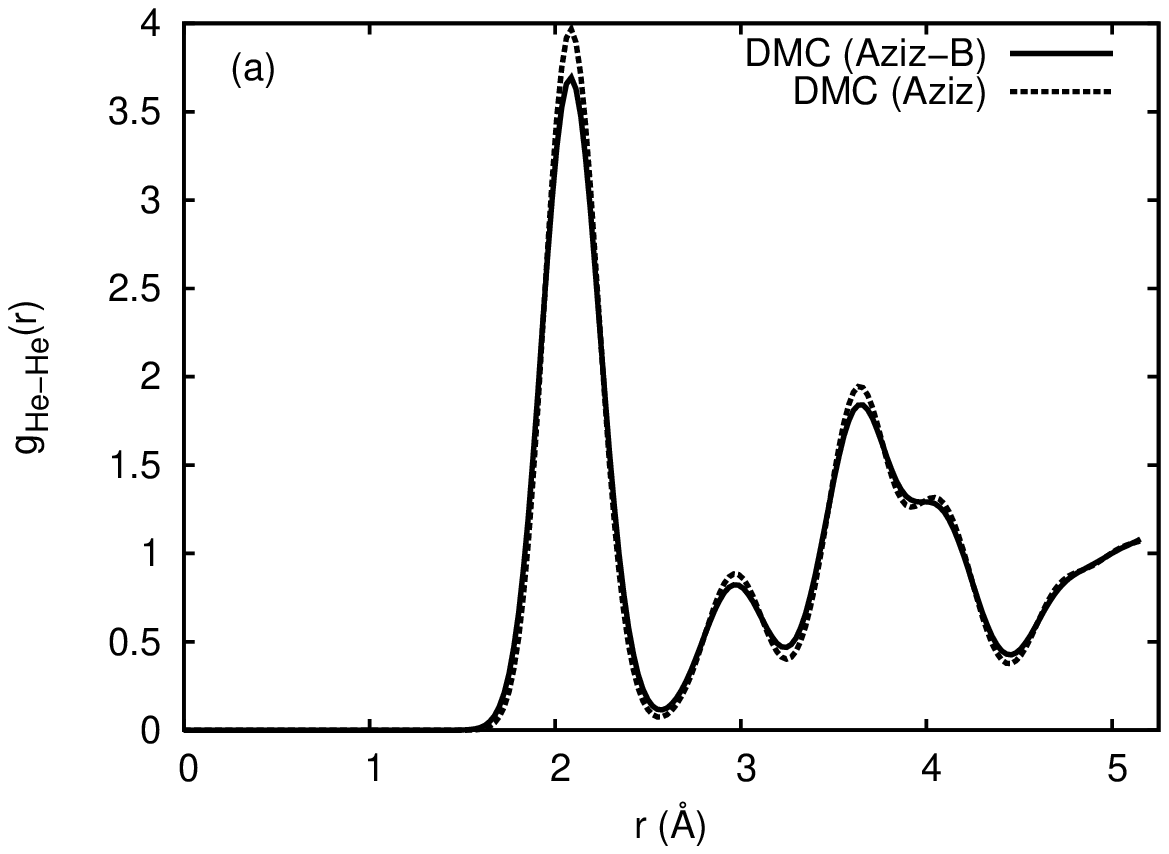}}
        {\includegraphics[width=1.0\linewidth]{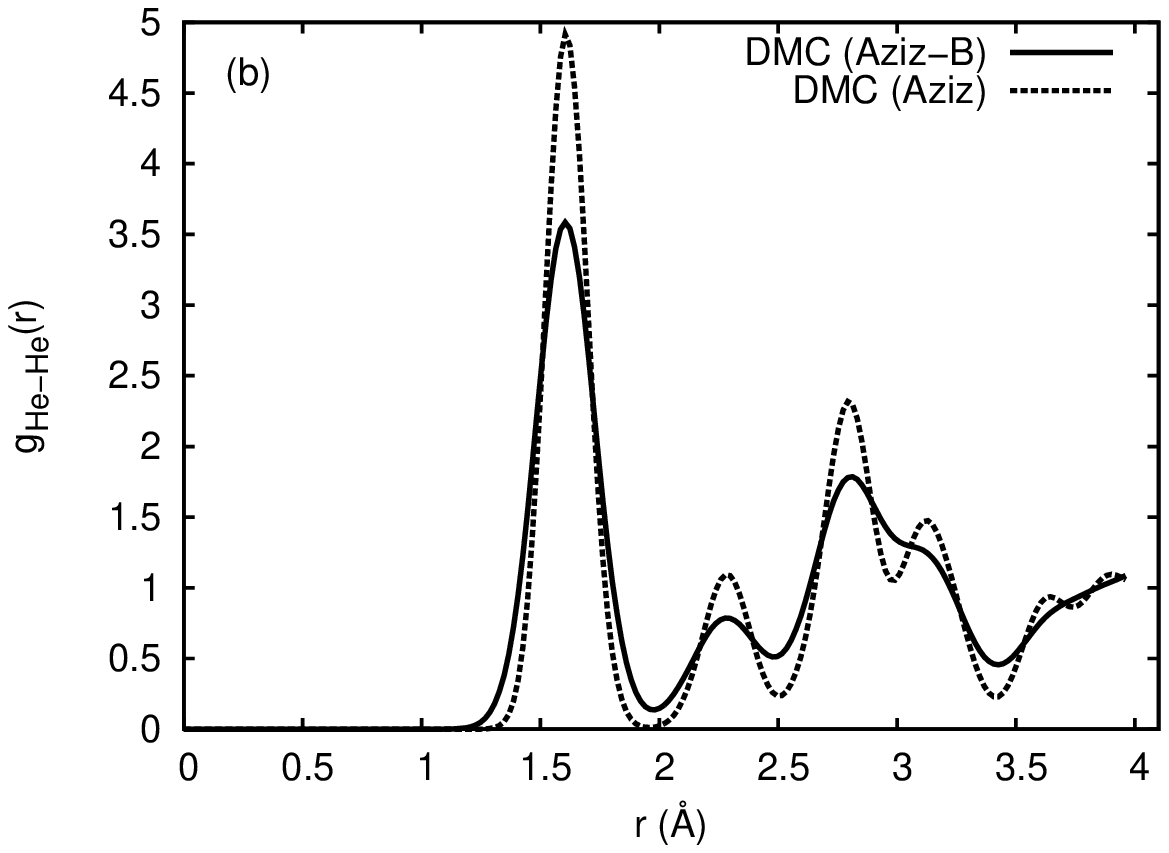}}
\caption{Radial pair distribution function calculated with the DMC method and 
         pure estimator techniques at volumes $V = 6.7$~\AA$^{3}$~(a) and 
         $3.0$~\AA$^{3}$~(b), employing the original and modified Aziz 
         pairwise potentials.}
\label{gr}
\end{figure}

In Fig.~\ref{profile}, we report the calculated radial distribution function
of $^{4}$He atoms, $\mu (r)$, around their equilibrium hcp lattice positions 
at two different volumes, using the DMC and pure estimator techniques.~\cite{barnett91,casulleras95}
In both cases, it is appreciated that the possibility of finding an atom at a
distance larger than $\sim 0.4$~\AA~ from its lattice site is practically
zero. As density is increased, the value of the $\mu (r)$ function at the 
origin increases noticeably whereas the variations on its tail turn out to
be less significant (see inset in Fig.~\ref{profile}). This finding is 
consistent with the $\gamma$ results explained above and illustrates 
the high degree of atomic delocalization in dense solid helium.   
In the same figure, we show Gaussian fits to the $\mu (r)$ results performed
in the radial distance interval $r \le 0.5$~\AA. It is found that these curves 
reproduce very well the computed $\mu (r)$ profiles (in fact, reduced chi-square 
values associated to our data fitting are close to unity). Also, we estimated 
the kurtosis in the three Cartesian directions (i.e., ${\bf \zeta} = \langle {\bf u}^{4} \rangle 
/ \langle {\bf u}^{2} \rangle^{2} - 3$)~\cite{cazorla08d} and found values  
compatible with zero in all the cases (i.e., $0.01-0.001$). 

Finally, in Fig.~\ref{gr} we enclose the radial pair distribution function calculated in 
dense $^{4}$He, considering both the Aziz-B and Aziz interaction models, with the 
DMC and pure estimator techniques.~\cite{barnett91,casulleras95} It is observed that the 
ground-state system rendered by the Aziz-B interaction is less structured than the one 
obtained with the original Aziz potential, due to its softer core. The 
position of the $g(r)$ maxima, however, 
roughly appear at the same distances in the two cases. It is also appreciated  
that, even at the smallest volume considered in this work (i.e., $V = 3.0$~\AA$^{3}$),
the minimum average distance between particles is larger than $d \sim 1.0$~\AA, 
that is, the threshold radius for the Aziz-B interaction model to be physically meaningful 
(see Sec.\ref{subsec:fitpot}).

\section{Conclusions}
\label{sec:conclusions}
We have performed a computational study of the quantum nuclear effects in compressed
$^{4}$He at zero temperature by relying exclusively on first-principles methods.  
For the description of the electronic degrees of freedom, we employ a 
non-standard implementation of density functional theory (DFT) which is able to 
deal efficiently with long-range van der Waals interactions. For the simulation 
of quantum nuclear effects, we employ the diffusion Monte Carlo method 
and a modified version of the pairwise Aziz potential, Aziz-B, that closely 
reproduces the static compression curve obtained with DFT. The Aziz-B 
potential is softer than Aziz one at short distances in a way which is rather 
similar to the behavior observed in molecular hydrogen.~\cite{moraldi12,boninsegni13} 
This softening of the potential wall is an effective pairwise approximation to 
many-body interaction terms which, according to our DFT results, are predominantly 
attractive.~\cite{loubeyre87} 
In fact, the Aziz-B interaction model introduced in this work may be used by 
others for the simulation of solid $^4$He at high pressures and low temperatures.    
We find that when solid helium is compressed the resulting gain in potential energy 
largely surpasses the accompanying increase in the kinetic energy. In particular, we 
show that the kinetic energy of $^{4}$He atoms increases very slightly under compression
at pressures larger than $\sim 85$~GPa. 
Also, we find that the Lindemann ratio in dense solid helium amounts to $0.10$ almost
independently of pressure. These results evidence the presence of strong quantum 
correlations in compressed $^{4}$He crystals, which allow the atoms to remain 
remarkably delocalized over a wide range of pressures. In addition to this, 
we perform analogous calculations using the quasi-harmonic DFT approach. 
We find that this method, which customarily is employed in 
computational high-$P$ studies, cannot reproduce with reliability the kinetic
energy and structural traits of compressed $^{4}$He at zero temperature. 
In particular, the kinetic energy discrepancies found with respect to the full 
quantum calculations amount to at least $50$~\%. The conclusions presented in this work 
are of critical importance for modeling of light and weakly interacting materials 
(e.g., H$_{2}$, CH$_{4}$, and NH$_{3}$) done in high-pressure studies and related to 
Earth and planetary sciences.

\begin{acknowledgments}
This research was supported under the Australian Research Council's
Future Fellowship funding scheme (project number RG134363), MICINN-Spain 
[Grants No. MAT2010-18113, CSD2007-00041, and FIS2011-25275], and Generalitat 
de Catalunya [Grant No.~2009SGR-1003]. 
\end{acknowledgments}

\end{document}